\documentclass[amsmath,amssymb,aps,prl,twocolumn,showpacs,floatfix]{revtex4}
\usepackage{graphicx}% Include figure files
\usepackage{dcolumn}% Align table columns on decimal point
\usepackage{bm}% bold math
\usepackage{color}
\newcommand{\trace}{{\rm Tr}}

\begin{document}

%\title{Nonequilibrium Thermodynamics of a Driven Quantum Level\\ Strongly Coupled to its Reservoirs}
%\title{Nonequilibrium Thermodynamics for Strongly Coupled Open Quantum Systems:\\ A Nonequilibrium Green Functions Approach}
\title{Quantum Thermodynamics: A Nonequilibrium Green's Functions Approach}

\author{Massimiliano Esposito}
\affiliation{Complex Systems and Statistical Mechanics, University of Luxembourg, L-1511 Luxembourg, Luxembourg}
\author{Maicol A. Ochoa}
\affiliation{Department of Chemistry \& Biochemistry, University of California San Diego, La Jolla CA 92093, USA}
\author{Michael Galperin}
\affiliation{Department of Chemistry \& Biochemistry, University of California San Diego, La Jolla CA 92093, USA}

\date{\today}

\begin{abstract}
We establish the foundations of a nonequilibrium theory of quantum thermodynamics 
for noninteracting open quantum systems strongly coupled to their reservoirs within 
the framework of the nonequilibrium Green functions (NEGF).
The energy of the system and its coupling to the reservoirs are controlled by a slow 
external time-dependent force treated to first order beyond the quasistatic limit. 
%We specifically consider a fermionic quantum level strongly 
%coupled to multiple reservoirs and subjected to a slow driving. 
We derive the four basic laws of thermodynamics and characterize 
reversible transformations. 
Stochastic thermodynamics is recovered in the weak coupling limit.
\end{abstract}

\pacs{
05.70.Ln,  %	Nonequilibrium and irreversible thermodynamics 
05.60.Gg,  %	Quantum transport 
05.70.-a   %    Thermodynamics
}

\maketitle

%%%%%%%%%%%%%%%%%%%%%%%%%%%%%%%%%%%%%%%%%%%%%%%%%%%%%%%%%%%%%%%%%%%%%%%%%%%%%%%%%%%%%%%%%%%%%%%%%%%%%%%%%%%%%%%%%%%%%%%%%%%%%%%%%%

Nonequilibrium thermodynamics of open quantum systems is a 
powerful tool for the study of mesoscopic and nanoscale systems.
It allows one to reliably assess the performance of energy-converting devices such as thermoelectrics or 
photoelectrics, by identifying the system entropy production. It enables one to meaningfully compare 
these different devices by discriminating the system-specific features from the universal ones
and to appraise the role of quantum effects. It can also be used to verify the thermodynamic 
consistency of approximation schemes.
Such a theory is nowadays available for systems weakly interacting with their surrounding 
\cite{SpohnLebowitz78, Breuer02, Esposito12, EspoStrassSchaBrandPRE13, Kosloff13, SchallerBook14} 
where it has proven very useful \cite{EspoRuttCleuPRB, EspoLindVdB_EPL09_Dot, EspoKumLindVdBPRE12, 
EspositoCuetaraGaspard11, SeifertBrandnerPRL13, SeifertBrandnerNJP13, KosloffFeldmannPRE12, 
AlickiKurizkiGelbPRE13, LutzSingerPRL12}.
However, in case of strong system-reservoir interactions, finding definitions for heat, 
work, entropy and entropy production, which satisfy the basic laws of thermodynamics is 
an open problem. 
Each proposal has its own limitations 
\cite{Jarzynski99, Allahverdyan00, AllahverdyanPRB, LutzPRA09, HanggiTalknerCampisiPRL09, 
EspoLindVdBNJP10, EspoPucciPeliti13, SanchezPRB14}, even at equilibrium 
\cite{FordOConnellPRL06, HanggiIngoldActaPol06, KimMahler07, 
HanggiIngoldTalkner08NJP, HanggiPRE09, CampisiTalknerCP10, KimMahlerPRE10}.
Reversible transformations, for instance, are never explicitly characterized.
Establishing a consistent nonequilibrium thermodynamics for open quantum systems 
strongly coupled to their surrounding is therefore an important step towards a 
more realistic thermodynamic description of mesoscopic and nanoscale devices.
It is also essential to improve our understanding of the microscopic foundations 
of thermodynamics.

In this Letter, we use the NEGF to establish a fully consistent nonequilibrium thermodynamic 
description of a fermionic single quantum level strongly coupled to multiple fermionic reservoirs.
A slow time-dependent driving force controls the level energy as well as the system-reservoir interaction.
We propose definitions for the particle number, the energy, and the entropy of the system, as well 
as for entropy production, heat and work, which give rise to a consistent zeroth, first, second, 
and third law. These definitions can be seen as energy resolved versions of the weak coupling 
definitions used in stochastic thermodynamics.
An interesting outcome of our approach is that the general form of the energy and particle 
currents are different from the standard form used in the NEGF and cannot be expressed as 
an expectation value of operators. 
We recover the known expressions when considering nonequilibrium steady 
states (i.e. in absence of driving) or in the weak coupling limit.

\begin{figure}[t]
\centering\rotatebox{0}{\scalebox{0.8}{\includegraphics[width=\linewidth]{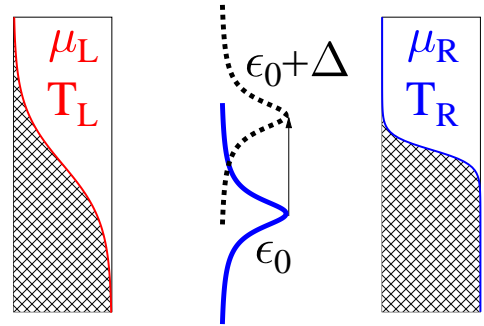}}}
\caption{\label{fig1}
(Color online) Sketch of a fermionic single quantum level junction. The level is 
broadened by the strong coupling to the reservoirs and is driven by a time-dependent force.
}
\end{figure}

The total Hamiltonian that we consider is 
$\hat H(t)=\hat H_S(t)+\sum_{\nu} \hat H_{\nu} + \sum_{\nu} \hat V_{\nu}(t)$, 
where $\nu$ labels the different fermionic reservoirs (see Fig.~\ref{fig1}), 
$\hat H_S(t) = \varepsilon (t) \hat d^\dagger \hat d$ is the fermionic single level Hamiltonian, 
$\hat H_{\nu}=\sum_{k\in\nu} \varepsilon_k \hat c_k^\dagger\hat c_k$ is the reservoir $\nu$ Hamiltonian, 
and $\hat V_{\nu}(t)=\sum_{k\in\nu}\left(V^\nu_{k}(t) \hat d^\dagger \hat c_k+\mbox{H.c.}\right)$
is the level-reservoir coupling. The time dependence in the system and in the coupling 
is due to the external time-dependent driving force.

%%%%%%%%%%%%%%%%%%%%%%%%%%%%%%%%%%%%%%%%%%%%%
%%%%%%%%%%%%%%%%%%%%%%%%%%%%%%%%%%%%%%%%%%%%%
 
The central object in the NEGF theory is the single particle Green function (GF)~\cite{HaugJauho_2008}
\begin{equation}\label{defG}
G(\tau_1,\tau_2) = -i\left\langle T_c\,\hat d(\tau_1)\,\hat d^\dagger(\tau_2) \right \rangle,
\end{equation}
where $T_c$ denotes the contour ordering operator and $\tau_1$ and $\tau_2$ are the contour variables. 
Here and below, $\hbar=k_B=1$.
When the time-dependent driving force is slow relative to the system relaxation 
time, the dynamics of the GF (\ref{defG}) can be evaluated using the first order 
gradient expansion \cite{LipavskySpickaVelickyPRB86, HaugJauho_2008, RammerBook2}.
Within this limit, this system dynamics is fully characterized by two quantities,
the probability to find the level filled at the energy $E$, $\phi(t,E)$, and the 
retarded projection of the Green function $G^r(t,E)$. 
%The variable $t =(t_1+t_2)/2$ characterizes the slow (classical) timescale, 
%where $t_1$ and $t_2$ are the physical times corresponding to the contour variables 
%$\tau_1$ and $\tau_2$ in Eq.~(\ref{defG}), and $E$ is the Fourier-conjugated 
%variable to the fast (quantum) timescale $t_q = t_1-t_2$.
The energy dependence of these quantities results from the fact that the energy 
of the level is not sharply defined at $\varepsilon(t)$ as in the weak coupling 
limit, but gets broadened by the strong coupling to the reservoirs.   
As shown is \cite{LipavskySpickaVelickyPRB86, BotermansMalflietPhysRep90, 
VoskresenskyNP00, KitaProgrTheorPhys10, VoskresenskyJPhysG13} (see also \cite{SupMat}),
the retarded Green function is given by 
\begin{align} \label{grad1Gr}
G^r(t,E)=\left[E-\varepsilon(t)-\Sigma^r(t,E)\right]^{-1},
\end{align}
where the real and imaginary part of the total retarded self-energy, $\Sigma^r(t,E)= \Lambda(t,E)-i\Gamma(t,E)/2$,
describe, respectively, the Lamb shift $\Lambda(t,E)$ and the broadening $\Gamma(t,E)$ of the system level caused 
by the coupling. In the weak coupling limit, $\Gamma \to 0$ and $\Lambda \to 0$.
The occupation probability of the level, $\phi(t,E)$, is obtained by solving the equation of motion
\begin{align}
\label{grad1Glt}
&\left\{E-\varepsilon(t)-\Lambda(t,E);A(t,E)\,\phi(t,E)\right\} \\ &\qquad
+\left\{\mbox{Re}\, G^r(t,E);\Gamma(t,E)\,\phi(t,E)\right\} =\mathcal{C}(t,E), \nonumber
\end{align}
where $\{f_1;f_2\}$ denotes the Poisson bracket operation $\partial_E f_1 \partial_t f_2-\partial_t f_1 \partial_E f_2$ 
and $A(t,E)= -2\, \mbox{Im} \, G^r(t,E)$ is the system spectral function describing
the Lorentzian probability amplitude for finding the system at energy $E$
\begin{align} \label{grad1A}
A(t,E)= \frac{\Gamma(t,E)}{\left(E-\varepsilon(t)-\Lambda(t,E)\right)^2 + \left(\Gamma(t,E)/2\right)^2}.
\end{align}
It become a delta function centered around $\varepsilon(t)$ in the weak coupling limit.
%The latter is the  first order gradient expansion of the sum of retarded self-energies for each reservoir
%$\Sigma_\nu^r(t,t')=\sum_{k\in\nu} V_k^\nu(t)\, g_k^r(t-t')  \,\overset{*}{V}{}_k^\nu(t')$,
%where $g_k^r(t-t')= -i\theta(t-t') e^{-i\varepsilon_k(t-t')}$ 
%is the retarded GF of free electrons in reservoir $\nu$. 
$\Sigma^r$ as well as $\Lambda$ and $\Gamma$ are sums of reservoirs contributions: 
respectively $\Sigma^r_\nu(t,E)$, $\Lambda_\nu(t,E)$ and $\Gamma_{\nu}(t,E)$.
Finally, the net particle current entering the level at energy $E$, $\mathcal{C}(t,E)$ in (\ref{grad1Glt}), 
is also the sum of different reservoirs contributions, each expressed as a difference between incoming (+) 
and outgoing (-) electrons  
%\begin{align}
%\label{defC}
%\mathcal{C}_\nu(t,E)=& A(t,E) \Gamma_\nu(t,E) \\ &\times
%\bigg(f_\nu(E)\left[1-\phi(t,E)\right]-\phi(t,E)\left[1-f_\nu(E)\right]\bigg) \nonumber,
%\end{align}
\begin{align}
\label{defC}
&\mathcal{C}_\nu(t,E) =\mathcal{C}_\nu^+(t,E) - \mathcal{C}_\nu^-(t,E) \\
&\mathcal{C}_\nu^+(t,E) = A(t,E) \Gamma_\nu(t,E) f_\nu(E)\left[1-\phi(t,E)\right] \nonumber \\
&\mathcal{C}_\nu^-(t,E) = A(t,E) \Gamma_\nu(t,E) \phi(t,E)\left[1-f_\nu(E)\right] \nonumber,
\end{align}
where $f_\nu(E)$ is the Fermi-Dirac distribution of reservoir $\nu$.

In absence of time-dependent driving, $\varepsilon$, $\Lambda$ and 
$\Gamma$ do not depend on time. If the level is in contact with a single 
reservoir at temperature $T$ and chemical potential $\mu$, it will relax 
to an equilibrium state where $\phi(t,E)$ will correspond to the Fermi 
distribution $f(E)$ at $T$ and $\mu$. If another reservoir at the same 
$T$ and $\mu$ is put in contact with the level, the system will remain 
at equilibrium with respect to the two reservoirs. In that sense, 
the NEGF satisfies the zeroth law of thermodynamics.

We introduce the renormalized spectral function
\begin{align}\label{defAnew} \hspace{-0.1cm}
\mathcal{A}(t,E) = A (1- \partial_E \Lambda) +\Gamma \partial_E \mbox{Re}G^r \geq 0,
\end{align}
which as its standard version (\ref{grad1A}), can be proven non-negative, normalized to one, and to 
converge to a delta in the weak coupling limit $\mathcal{A} \to 2 \pi \delta(E-\varepsilon) $\cite{SupMat}.
We define the particle number, energy and entropy of the system as energy-resolved 
versions of the standard weak coupling definitions where the energy resolution is 
controlled by the renormalized spectral function $\mathcal{A}$
\begin{align}
\mathcal{N}(t) =& \int\frac{dE}{2\pi}\, \mathcal{A}(t,E)\, \phi(t,E) \label{particlenb} \\
\mathcal{E}(t) =& \int\frac{dE}{2\pi}\, \mathcal{A}(t,E)\, E\, \phi(t,E) \label{energy} \\
\mathcal{S}(t) =& \int\frac{dE}{2\pi}\, \mathcal{A}(t,E)\, \sigma(t,E), \label{entropy} 
\end{align}
where $\sigma(t,E)$ is an energy resolved Shannon entropy
\begin{align} \label{EresEntropy}
\sigma(t,E) = & -\phi(t,E) \ln\phi(t,E) \nonumber\\
& -[1-\phi(t,E)]\ln[1-\phi(t,E)].
\end{align}
When attempting to use the standard spectral function rather then the renormalized one in 
(\ref{particlenb})-(\ref{entropy}), one fails to define a proper entropy production and second law. 

The entropy (\ref{entropy}) was introduced in Refs.~\cite{VoskresenskyNP00, 
KitaProgrTheorPhys10} in the context of the quantum Boltzmann equation. 
We emphasize that this entropy satisfies the third law. Indeed at equilibrium when $\phi(E)=f(E)$, 
if we take the limit $T \to 0$, $\sigma^{eq}(E) \to 0$ and therefore $\mathcal{S}^{eq} \to 0$.

The evolution of the particle number (\ref{particlenb}) 
\begin{equation} \label{PartCons}
d_t \mathcal{N}(t) = \sum_{\nu} \mathcal{I}_{\nu}(t)
\end{equation}
is given by the sum of the energy-integrated particle current (\ref{defC}) from reservoir $\nu$
\begin{equation} \label{PartCurr}
\mathcal{I}_{\nu}(t) =\int\frac{dE}{2\pi}\, \mathcal{C}_\nu(t,E).
\end{equation}

The evolution of the energy (\ref{energy}) in turn can be expressed as a first law
\begin{equation} \label{1stlaw}
d_t \mathcal{E}(t) = \sum_\nu \dot{\mathcal{Q}}_\nu(t) + \dot{\mathcal{W}} + \dot{\mathcal{W}}_{c}.
\end{equation}
Note that the dots are not partial derivatives, but a symbolic notation for rates. 
The first contribution is the heat flux from reservoir $\nu$ 
\begin{align} \label{defQnu}
\dot{\mathcal{Q}}_\nu = \mathcal{J}_{\nu}(t) - \mu_\nu\, \mathcal{I}_{\nu}(t),
\end{align}
where the energy current from reservoir $\nu$ is the energy integral 
of the energy times the particle current (\ref{defC}) at that energy
\begin{equation} \label{EnCurrJ}
\mathcal{J}_{\nu}(t) = \int\frac{dE}{2\pi}\, E\, \mathcal{C}_\nu(t,E).
\end{equation}
The second is the mechanical work performed by the external time-dependent force 
\begin{align}
\label{defW}
\dot{\mathcal{W}}(t) = \int\frac{dE}{2\pi} \big(
- A \,\phi\, \partial_t (E-\varepsilon(t)-\Lambda) 
- \Gamma\, \phi\, \partial_t \mbox{Re}\, G^r \big)
\end{align}
and the third is the chemical work due to the particle currents flowing from the reservoirs to the system
\begin{equation} \label{ChemWork}
\dot{\mathcal{W}}_{c} = \sum_{\nu} \mu_\nu\, \mathcal{I}_{\nu}(t).
\end{equation}

The evolution of the entropy (\ref{entropy}) can be expressed as a second law
\begin{equation} \label{2ndlaw}
d_t \mathcal{S}(t) = \dot{\mathcal{S}}_i(t) + \sum_\nu \frac{\dot{\mathcal{Q}}_\nu(t)}{T_\nu},
\end{equation}
where the entropy production becomes an energy-resolved version of the weak coupling form  
%\begin{align}\label{EPdef} \hspace{-0.2cm}
%\dot{\mathcal{S}}_i(t) =
%\sum_\nu \int \frac{dE}{2\pi}\, \mathcal{C}_\nu(t,E)\ln\frac{[1-\phi(t,E)]f_\nu(E)}{\phi(t,E)[1-f_\nu(E)]} \geq 0 
%\end{align}
\begin{align}\label{EPdef} 
\dot{\mathcal{S}}_i(t) =
\sum_\nu \int \frac{dE}{2\pi}\, \big( \mathcal{C}_\nu^+(t,E)-\mathcal{C}_\nu^-(t,E) \big) 
\ln\frac{\mathcal{C}_\nu^+(t,E)}{\mathcal{C}_\nu^-(t,E)} \geq 0
\end{align}
which measures the deviation from detailed balance at each energy $E$,
and only vanishes at equilibrium when $\forall \nu: f_\nu(E)=\phi(t,E)$.

In the presence of a single reservoir, the second law (\ref{2ndlaw}) implies $\dot{\mathcal{Q}} \leq T d_t \mathcal{S}(t)$.
When integrated along transformations connecting an initial and final equilibrium point
we recover Clausius inequality $\mathcal{Q} \leq T \Delta \mathcal{S}^{eq}$. 
Introducing the nonequilibrium grand potential
\begin{align}
\Omega(t) = \mathcal{E}(t) - \mu\, \mathcal{N}(t) - T \mathcal{S}(t) \label{GrandPot}
\end{align}
and using the first law (\ref{1stlaw}), the second law (\ref{2ndlaw}) can also be rewritten as
\begin{align}\label{2ndlawBis}
T \dot{\mathcal{S}}_i(t) = \dot{\mathcal{W}}(t) - d_t \Omega(t) \geq 0.
\end{align}
For a transformation starting and ending at equilibrium, we thus recover Kelvin's statement of the second law 
$\mathcal{W}(t) \geq \Delta \Omega^{eq}$, where $\Omega^{eq}=T \int\frac{dE}{2\pi}\, \mathcal{A}(t,E)\, \ln f(E)$.

For reversible transformations, the inequalities resulting from the positivity of the entropy production become equalities.
Such transformation occurs when the level is in contact with a single reservoir and subjected to a quasistatic driving 
(much slower than the level relaxation time).
In this case, the entropy production vanishes to first order $\dot{\mathcal{S}}_i(t)=0$, while to the same order 
heat and mechanical work become state functions $\dot{\mathcal{Q}}(t)/T = d_t \mathcal{S}^{eq}$ 
and $\dot{\mathcal{W}}(t) = d_t \Omega^{eq}$.

We can also prove that (as for weak coupling \cite{EspoVdB_EPL_11}) the nonequilibrium 
grand potential is always larger then the equilibrium one, i.e. $\Omega(t) \geq \Omega^{eq}$.
Indeed, using (\ref{GrandPot}) and (\ref{particlenb})-(\ref{entropy}), we find that
\begin{align}\label{IneqOmega}
\Omega(t)-\Omega^{eq} = T \int\frac{dE}{2\pi}\, \mathcal{A}(t,E)\, D(t,E) \geq 0 ,
\end{align}
where the energy-resolved relative entropy reads
\begin{align}\label{RelEnt}
D(t,E) =& \; \phi(t,E) \ln \frac{\phi(t,E)}{f(E)} \nonumber\\
& +\left[1-\phi(t,E)\right]\ln\frac{1-\phi(t,E)}{1-f(E)} \geq 0.
\end{align}
The non-negativity of (\ref{IneqOmega}) follows from $\mathcal{A},D \geq 0$.

%%%%%%%%%%%%%%%%%%%%%%%%%%%%%%%%%%%%%%%%%%%%%
\begin{figure}[t]
\centering\includegraphics[width=\linewidth]{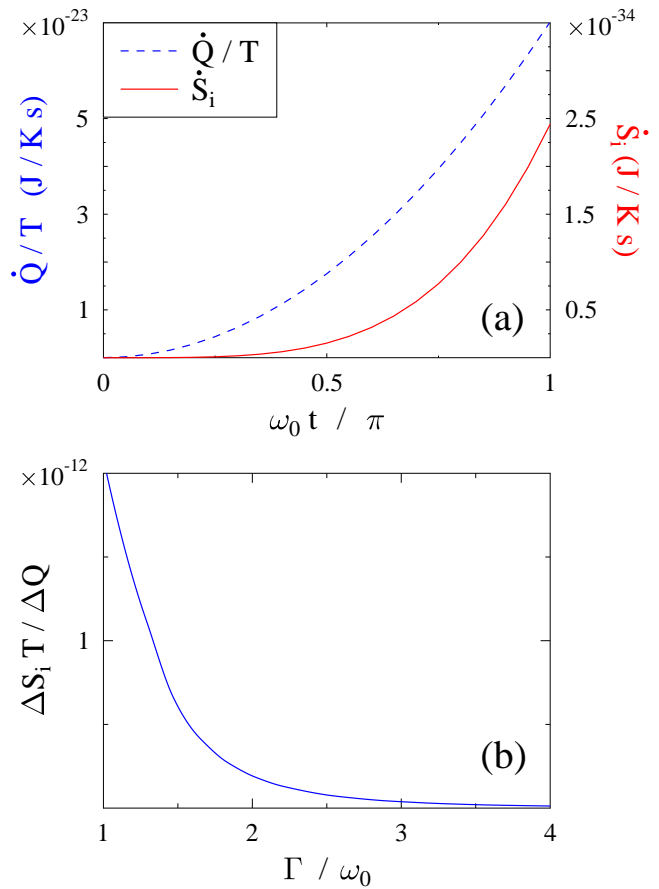}
\caption{\label{fig2}
(Color online) Heat flux (\ref{defQnu}) and entropy production (\ref{EPdef}), for the 
quantum level in contact with a single reservoir at $T=300$~K. The external force drives 
the level energy as $\varepsilon(t)=\varepsilon_0+\Delta\left(1-\cos\omega_0 t\right)/2$ 
from $\varepsilon_0$ at $t=0$ to $\varepsilon_0+\Delta$ at $t=\pi/\omega_0$, where 
$\varepsilon_0=-0.02$~eV and $\Delta=0.02$~eV.
Entropy production (solid line, red) and heat flux (dashed line, blue) are depicted in (a) as functions of time for $\Gamma=0.01$~eV and $\omega_0=0.01$~eV. 
The ratio of their time-integrated values is depicted in (b) as a function of the driving rate $\omega_0$. 
}
\end{figure}
%%%%%%%%%%%%%%%%%%%%%%%%%%%%%%%%%%%%%%%%%%%%%

We consider in Fig.~\ref{fig2} the quantum level in contact with a single reservoir. 
Its energy is driven by the external force according to the protocol described in the caption. 
Fig.~\ref{fig2}a depicts the heat flux (\ref{defQnu}) and entropy production (\ref{EPdef}) 
increase with time as the distribution $\phi$ departs from its equilibrium value.
The reversible transformation ($\dot{S}_i=0$) is reached in the very slow driving limit when 
$\omega_0 \to 0$, as shown on Fig.~\ref{fig2}b.

We note that the system energy (\ref{energy}) and particle number (\ref{particlenb}) 
as well as the energy and particle currents (\ref{EnCurrJ}) and (\ref{PartCurr}) that 
we introduced cannot be expressed in term of expectation values of operators. 
One may interpret this as a manifestation of the fact that defining a boundary between 
the system and the reservoirs in case of strong interaction is an ambiguous task.
The main argument in favor of the proposed definitions is that they 
lead to a consistent nonequilibrium thermodynamics at slow driving. 

In absence of driving the system eventually reaches a steady state (equilibrium or nonequilibrium),
where the system properties such as $\phi(t,E)$, $\mathcal{A}(t,E)$ and 
(\ref{particlenb})-(\ref{entropy}) become time independent. In this case we find that
$\mathcal{J}_{\nu}(t)= -\trace \big( \hat H_\nu\, d_t \hat \rho(t) \big)$ and 
$\mathcal{I}_{\nu}(t)=- \sum_{k\in {\nu}} \trace \big( \hat c^\dagger_{k} \hat c_{k}\, d_t \hat \rho(t) \big)$ \cite{SupMat}.
The first and second law at steady state simplify to
\begin{align} \label{2ndlaw_nodrv}
\dot{\mathcal{W}}_{c} = - \sum_\nu \dot{\mathcal{Q}}_\nu(t), \ \ 
\dot{\mathcal{S}}_i(t) =- \sum_\nu \frac{\dot{\mathcal{Q}}_\nu(t)}{T_\nu} \geq 0.
\end{align}

Since in the weak coupling limit $A$ and $\mathcal{A}$ become delta functions, 
we recover the usual definitions of stochastic thermodynamics 
\cite{Seifert12Rev, Esposito12, EspVDBRev2014} for a master equation 
with Fermi's golden rule rates describing the evolution of the occupation probability 
of the level \cite{EspoLindVdB_EPL09_Dot, EspKawLindVdBEPL10}.

%%%%%%%%%%%%%%%%%%%%%%%%%%%%%%%%%%%%%%%%%%%%%
\begin{figure}[t]
\centering\includegraphics[width=\linewidth]{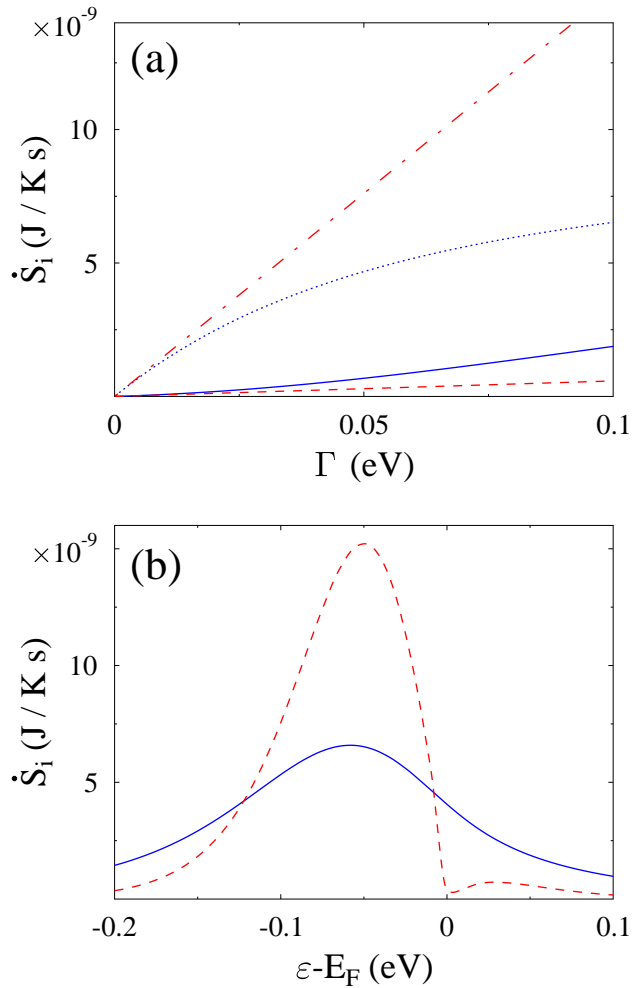}
\caption{\label{fig3}
(Color online) Entropy production for the quantum level at steady state between two reservoirs 
$\nu=L,R$ (with $T_L=300$~K, $T_R=10$~K, $\mu_L=-0.05$~eV, $\mu_R=E_F=0$) as a function of: 
(a) the interaction strength to the reservoirs $\Gamma=2\Gamma_L=2\Gamma_R$,
(b) the position of the level $\varepsilon$. 
The strong coupling entropy production (\ref{EPdef}) is depicted for $\varepsilon=-0.05$~eV (dotted line, blue) 
and $\varepsilon=0.05$~eV (solid line, blue) and its weak coupling counterpart for $\varepsilon=-0.05$~eV 
(dash-dotted line, red) and $\varepsilon=0.05$~eV (dashed line, red).
The energy grid used spans from $-3$~eV to $3$~eV with step $10^{-6}$~eV.
}
\end{figure}
%%%%%%%%%%%%%%%%%%%%%%%%%%%%%%%%%%%%%%%%%%%%%

Figure~\ref{fig3}a depicts the entropy production of the quantum level at steady 
state between two reservoirs with different temperatures and chemical potentials. 
The entropy production is plotted as a function of the coupling strength 
with the reservoirs, when this device operates as a thermoelectric.
As the coupling strength increases, the discrepancy between the entropy 
production (\ref{EPdef}) and its weak coupling counterpart (dotted vs. 
dash-dotted and solid vs. dashed lines) becomes more pronounced.
We note that the weak coupling prediction can overestimate (dash-dotted line) 
or underestimate (dashed line) the entropy production (\ref{EPdef}).
Fig.~\ref{fig3}b depicts the same two entropy productions for 
$\Gamma=0.1$~eV as functions of the position of the level.
In the weak coupling regime, this system satisfies the condition of tight 
coupling (energy and particle current are proportional) 
\cite{LinkePRL02, EspositoPRL09, IrmyPRE14, WhitneyPRL14} which enables it 
to operate reversibly at finite bias, as seen at $\varepsilon=0.0017$~eV. 
However, the level broadening induced by the strong coupling to the 
reservoirs completely breaks the tight coupling property and reversibility is lost.

The main message of this Letter is that it is possible to formulate a consistent nonequilibrium 
thermodynamics for driven open quantum systems strongly coupled to their reservoirs. 
No such theory existed before and the definitions we used seem to be the only ones 
rendering such a formulation possible.
We considered a fermionic level coupled to fermionic reservoirs, but 
our approach can be straightforwardly extended to any noninteracting fermionic 
or bosonics systems. It can probably be extended to describe interacting 
systems, but considering fast drivings remains out of its reach since it relies 
on treating slow time-dependent driving forces (i.e. gradient expansion).
We are now in the position to address important problems such as characterizing 
the dissipation caused by connecting or disconnecting a system from its reservoirs, 
or assessing the difference in performance between strongly-coupled 
and weakly coupled energy converting devices.

%%%%%%%%%%%%%%%%%%%%%%%%%%%%%%%%%%%%%%%%%%%%%%%%%%%%%%%%%%%%%%%%%%%%%%
\begin{acknowledgments}
M.E. is supported by the National Research Fund, Luxembourg in the frame of project FNR/A11/02.
M.G. gratefully acknowledges support by the Department of Energy (Early Career Award, DE-SC0006422).
\end{acknowledgments}
%%%%%%%%%%%%%%%%%%%%%%%%%%%%%%%%%%%%%%%%%%%%%%%%%%%%%%%%%%%%%%%%%%%%%%%%%%%%%%%%%%%%%%%%%%%%%%%%%%%%%%%%%%%%%%%%%%%%%%%%
\section{Supplementary Material}
%%%%%%%%%%%%%%%%%%%%%%%%%%%%%%%%%%%%%%%%%%%%%%%%%%%%%%%%%%%%%%%%%%%%%%%%%%%%%%%%%%%
\subsection{The model}

We consider a single quantum level bi-linearly coupled to a set of Fermionic 
reservoirs $\{\nu\}$ (each at equilibrium with its own chemical potential $\mu_\nu$ 
and temperature $T_\nu$). The level energy as well as the couplings to the reservoirs
are controlled by an external time-dependent force. 
The Hamiltonian of the model is
\begin{equation}
 \hat H(t)=\hat H_S(t) + \sum_\nu\left( \hat H_\nu + \hat V_\nu(t)\right),
\end{equation}
where 
\begin{align}
&\hat H_S(t) = \varepsilon(t)\hat d^\dagger\hat d \\
&\hat H_\nu = \sum_{k\in\nu} \varepsilon_k\hat c_k^\dagger\hat c_k \\
&\hat V_\nu(t) = \sum_{k\in\nu}\left( V^\nu_{k}(t)\hat d_m^\dagger\hat c_k+\mbox{H.c.}\right)
\end{align}
represent respectively the single quantum level, the reservoir $\nu$, and the system-reservoir coupling.
$\hat d^\dagger$ ($\hat d$) and $\hat c_k^\dagger$ ($\hat c_k$) create (annihilate) 
an electron in the system level and in state $k$ of the reservoir, respectively. 
Here and below $\hbar=e=k_b=1$.

Following Ref.~\cite{JauhoWingreenMeirPRB94}
we assume that driving in the system-reservoirs couplings has the form
\begin{equation}
 V_{k}^\nu(t) = u_\nu(t)\, V^\nu_{k}\qquad u_\nu(t)\in \mathbb{R}.
\end{equation}

%%%%%%%%%%%%%%%%%%%%%%%%%%%%%%%%%%%%%%%%%%%%%%%%%%%%%%%%%%%%%%%%%%%%%%%%%%%%%%%%%%%
\subsection{Green function equation of motion}

The central object in nonequilibrium Green function (NEGF) theory is the 
single-particle GF defined on the Keldysh contour in Eq.~(1) of the letter ~\cite{HaugJauho_2008}.
Labeling the contour branches as time ordered, $s={}-{}$, and anti-time ordered, $s={}+{}$, 
one defines the four projections of the GF according to the position 
of the contour variables on the branches: $--$, $-+$, $+-$, and $++$.
Since the GF is a two-time correlation function~\footnote{Here and below we utilize Greek 
symbol $\tau$ for the contour variables, while the corresponding real times are denoted by 
their Latin analog $t$.}, one can write their equation of motion (EOM) in terms of the left 
or the right time. 
The EOMs for the projections $s_1s_2$ of the GF are given by~\cite{WagnerPRB91,HaugJauho_2008}
\begin{align}
 \label{LEOM}
 &\left(i\overset{\rightarrow}{\partial}{}_{t_1}-\varepsilon(t_1)\right)G^{s_1s_2}(t_1,t_2)
 = \sigma^z_{s_1s_2}\delta(t_1-t_2)
 \\ &\qquad
  + \sum_{s_3=-,+}\int dt_3\,\Sigma^{s_1s_3}(t_1,t_3)\,
\bigg( s_3\, G^{s_3s_2}(t_3,t_2) \bigg)
\nonumber \\
 \label{REOM}
 &G^{s_1s_2}(t_1,t_2)\left(-i\overset{\leftarrow}{\partial}_{t_2}-\varepsilon(t_2)\right)
 = \sigma^z_{s_1s_2}\delta(t_1-t_2) 
 \\ & \qquad
 + \sum_{s_3=-,+}\int dt_3\, G^{s_1s_3}(t_1,t_3)\, 
 \bigg( s_3\, \Sigma^{s_3s_2}(t_3,t_2) \bigg),
 \nonumber
\end{align}
where $\mathbf{\sigma}^z$ is the Pauli matrix, the arrows above the time derivatives
show the direction in which the operator acts, and $\Sigma^{s_1s_2}(t_1,t_2)$ 
is the $s_1s_2$ projection of the self-energy (SE) due to the system-reservoirs couplings
\begin{align}
\label{defSigma}
\Sigma(\tau_1,\tau_2) =& \sum_{\nu=L,R} \Sigma_\nu(\tau_1,\tau_2) \\
=& \sum_{\nu=L,R} \sum_{k\in \nu} V^\nu_{k}(t_1)\,g_k(\tau_1,\tau_2)\, \overset{*}{V}{}^\nu_{k}(t_2).
 \nonumber
\end{align}
Here 
\begin{equation}
\label{defgk}
g_k(\tau_1,\tau_2)=-i\langle T_c\,\hat c_k(\tau_1)\,\hat c_k^\dagger(\tau_2)\rangle
\end{equation}
is the Green function of free electrons in the reservoirs. 

We are going to use the retarded ($r$) and lesser ($<$) projections of the GF and SE. 
They are related to the $s_1s_2$ projections by
\begin{align}
G^{<}(t_1,t_2) =& G^{-+}(t_1,t_2) \\
G^r(t_1,t_2) =& G^{--}(t_1,t_2) + G^{-+}(t_1,t_2) .
\end{align}

%%%%%%%%%%%%%%%%%%%%%%%%%%%%%%%%%%%%%%%%%%%%%%%%%%%%%%%%%%%%%%%%%%%%%%%%%%%%%%%%%%%
\subsection{Gradient expansion}

The gradient expansion of a two-time function, $F(t_1,t_2)$, starts by introducing 
a slow classical, $t=(t_1+t_2)/2$, and a fast quantum, $t_q=t_1-t_2$, timescales,
and performing a Fourier transform of the fast time $t_q$
\begin{equation}
 F(t_1,t_2) = F(t,t_q) \to F(t,E)\equiv \int dt_q\, e^{iEt_q} F(t,t_q).
\end{equation}
A similar transformation for an integral expression of the form 
\begin{equation}
F(t_1,t_2) = \int dt_3\, F_1(t_1,t_3)\, F_2(t_3,t_2),
\end{equation}
appearing in the GF EOMs (\ref{LEOM}) and (\ref{REOM}), leads to \cite{HaugJauho_2008}
\begin{equation}
F(t,E) = F_1(t,E) \, \hat{\mathcal{G}}(t,E)\, F_2(t,E),
\end{equation}
where
\begin{equation}
 \label{gradop}
 \hat{\mathcal{G}}(t,E)=  \mbox{exp}\left(\frac{1}{2i}\left[
\overset{\leftarrow}{\partial}_{t}\,\overset{\rightarrow}{\partial}_{E} -
\overset{\leftarrow}{\partial}_{E}\,\overset{\rightarrow}{\partial}_{t}
 \right]\right)
\end{equation}
is the gradient operator. 

When the driving is slow relative to characteristic timescales of the system, 
one can expand (\ref{gradop}) in Taylor series and only keep the first two terms: 
\begin{equation}
\label{grad1}
F(t,E)\approx F_1(t,E)\, F_2(t,E)+\frac{i}{2}\left\{F_1(t,E);F_2(t,E)\right\},
\end{equation}
where
\begin{align}
& \left\{F_1(t,E);F_2(t,E)\right\}= 
 \\ &\qquad
\partial_E F_1(t,E)\cdot\partial_t F_2(t,E)
-\partial_t F_1(t,E)\cdot\partial_E F_2(t,E)
\nonumber
\end{align}
denotes the Poisson bracket. 
This is the essence of the gradient expansion technique.

%%%%%%%%%%%%%%%%%%%%%%%%%%%%%%%%%%%%%%%%%%%%%%%%%%%%%%%%%%%%%%%%%%%%%%%%%%%%%%%%%%%
\subsection{Slow driving of a quantum level strongly coupled to the reservoirs}

The couplings to the reservoirs induce a shift in the level position (i.e. the Lamb shift) as
well as its hybridization with the states of the reservoir (i.e. broadening). These effects 
are quantified by the real and imaginary parts of the retarded projection of the SE (\ref{defSigma})
\begin{align}
\Sigma^r(t_1,t_2)
=& \sum_{\nu=L,R} \sum_{k\in \nu} V^\nu_{k}(t_1)\,g_k^r(t_1,t_2)\, \overset{*}{V}{}^\nu_{k}(t_2) \\
=& \Lambda(t_1,t_2) -\frac{i}{2}\Gamma(t_1,t_2) \nonumber \\
g_k^r(t_1,t_2) =& -i\theta(t_1-t_2)\, e^{-i\varepsilon_k(t_1-t_2)}.
%\label{defLam}
%\Lambda(t_1,t_2) =& \mbox{Re}\, \Sigma^r(t_1,t_2) \\
%\label{defGam}
%\Gamma(t_1,t_2) =& -2\,\mbox{Im}\, \Sigma^r(t_1,t_2).
\end{align}
$\Gamma$ also characterizes the electronic escape rate and
its inverse measures the characteristic response time of the system. 

We now assume that the position of the level $\varepsilon(t)$ and its 
coupling to the reservoir $V^\nu_k(t)$ are driven by an external 
field which is slow compared to the characteristic time of the system.
Using the first order gradient expansion (\ref{grad1}) in  
(\ref{LEOM}) and (\ref{REOM}) one get EOMs (see Eqs.~(2) and (3) of the letter) 
for the first order gradient expansion of the retarded and lesser projections of the GF
\begin{align}
G^r(t_1,t_2) \to &\, G^r(t,E) \\
G^{<}(t_1,t_2) \to &\, i A(t,E)\,\phi(t,E),
\end{align}
where $G^r(t,E)$, $\phi(t,E)$ and $A(t,E)$ are given, respectively, 
by Eqs.~(2), (3) and (4) of the letter.

%%%%%%%%%%%%%%%%%%%%%%%%%%%%%%%%%%%%%%%%%%%%%%%%%%%%%%%%%%%%%%%%%%%%%%%%%%%%%%%%%%%
\subsection{Entropy balance}

The entropy of an open system may change either due to the (non-negative) 
entropy production, $\dot{\mathcal{S}}_i(t)$, or due to heat fluxes through 
the system-reservoirs boundaries $\{\nu\}$, $\dot{\mathcal{Q}}_\nu(t)/T_\nu$. 
This statement constitutes the nonequilibrium second law of thermodynamics 
detailed in Eq.~(18) of the letter.

Following Refs.~\cite{IvanovKnollVoskresenskyNuclPhysA99,KitaProgrTheorPhys10},
we multiply Eq.~(3) of the letter by $\ln\,[(1-\phi(t,E))/\phi(t,E)]$ and integrate 
the result over energy. Utilizing ($F_{1,2}$ are arbitrary functions of $t$ and $E$; 
we drop the $(t,E)$ dependence to shorten the notation) 
\begin{align}
 &\{F_1;F_2\sigma\}=
 -\{F_1;F_2\,\phi\}\ln\phi 
 \\ & \qquad\qquad\quad\,\,
 - \{F_1;F_2[1-\phi]\}\ln[1-\phi] 
 \nonumber 
\\
 & d\sigma = \ln\frac{1-\phi}{\phi}\, d\phi
 \\
 &\{E-\varepsilon(t)-\Lambda;A\}+\{\mbox{Re}\, G^r; \Gamma\} = 0,
\end{align}
where $\sigma(t,E)$ is the energy-resolved Shannon entropy given in Eq.~(10) 
of the letter, we find after lengthy but straightforward calculations that
\begin{equation}
\label{step1}
\frac{d\mathcal{S}(t)}{dt} = \int\frac{dE}{2\pi}\mathcal{C}(t,E)\ln\frac{1-\phi(t,E)}{\phi(t,E)}
\end{equation} 
with the energy-resolved current $\mathcal{C}(t,E)$ and entropy of the system 
$\mathcal{S}(t)$ defined in Eqs.~(5) and (9) of the letter, respectively.
Rewriting the right side of Eq.~(\ref{step1}) as
\begin{align}
& \sum_\nu \int\frac{dE}{2\pi} \mathcal{C}_\nu(t,E)
\ln \frac{[1-\phi(t,E)]\, f_\nu(E)}{\phi(t,E)\, [1-f_\nu(E)]}
\\ &
+\sum_\nu\int\frac{dE}{2\pi}\frac{E-\mu_\nu}{T_\nu}\mathcal{C}_\nu(t,E),
\nonumber
\end{align}
where we used $\ln\big([1-f_\nu(E)]/f_\nu(E)\big)=(E-\mu_\nu)/T_\nu$,
and identifying the two terms respectively as entropy production (see Eq.~(19) 
of the letter) and heat flux (see Eqs.~(12), (14) and (15) of the letter), we get the second 
law given by Eq.~(18) of the letter.

%%%%%%%%%%%%%%%%%%%%%%%%%%%%%%%%%%%%%%%%%%%%%%%%%%%%%%%%%%%%%%%%%%%%%%%%%%%%%%%%%%%
\subsection{Renormalized spectral function}

We note that satisfying the second law at an arbitrary (slow) driving is only 
possible when using the renormalized version of the spectral function
$\mathcal{A}(t,E)$ given by Eq.(6) of the letter. 
Similarly to the standard definition given by Eq.~(4) of the letter, 
the renormalized spectral function is non-negative and normalized.
Indeed, from its definition, it is easy to show that \cite{WeinholdFrimanNorenbergPhysLettB98, VoskresenskyNP00}
\begin{equation}
\mathcal{A}(t,E) = A(t,E) \Gamma(t,E) B(t,E) /2
\end{equation}
where
\begin{align} 
B(t,E)=& -2\,\mbox{Im}\big[\big(1-\partial_E \Sigma^r(t,E)\big) G^r(t,E)\big] \\
=& 2\,\mbox{Im}\, \partial_E \ln G^r(t,E).
\nonumber
\end{align}
Since for a non-interacting level \cite{Hewson1993}
\begin{equation}
B(t,E)=-2\,\mbox{Im} G^r(t,E)= A(t,E) \geq 0
\end{equation}
and since $A(t,E)$ and $\Gamma(t,E)$ are non-negative (see e.g. \cite{VoskresenskyJPhysG13}), 
we proved that $\mathcal{A}(t,E) \geq 0$.
Its normalization follows from 
\begin{align}
&\int\frac{dE}{2\pi} \mathcal{A}(t,E) = \nonumber \\
&\int\frac{dE}{2\pi}\bigg( A(t,E) -2\,\mbox{Im} G^r(t,E)\frac{\partial\Sigma^r(t,E)}{\partial E} \bigg) = \\ 
&\int\frac{dE}{2\pi} A(t,E) = 1 ,
\nonumber
\end{align}
where we used complex plane integration taking into account the fact that the 
poles of retarded Green function are in the lower half of the complex plane.

In the weak coupling limit, when $\Gamma\to 0$ and $\Lambda\to 0$, 
 $\mathcal{A} \to 2\pi\delta(E-\varepsilon)$.
Indeed, since $\mathcal{A}=A^2\,\Gamma/2$, then for an arbitrary function 
$\varphi(E)$ we find that
\begin{align*}
&\int_{-\infty}^{+\infty} dE\, \varphi(E)\, \mathcal{A}(E,t)\overset{\Gamma x=E-\varepsilon(t)}{=}
\int_{-\infty}^{+\infty} dx \frac{\varphi(\Gamma x+\varepsilon(t))}{2\left(x^2+1/4\right)^2}
\\ &\hspace{0.6cm}
\overset{\Gamma\to 0}{\longrightarrow} 
\frac{\varphi(\varepsilon(t))}{2}\int_{-\infty}^{+\infty} dx\frac{1}{\left(x^2+1/4\right)^2}
= 2\,\pi\, \varphi(\varepsilon(t)) .
\end{align*}

%%%%%%%%%%%%%%%%%%%%%%%%%%%%%%%%%%%%%%%%%%%%%%%%%%%%%%%%%%%%%%%%%%%%%%%%%%%%%%%%%%%
\subsection{Connection to previous definitions}

We note that if the renormalized spectral function (see Eq.~(6) of the letter) 
is replaced by hand by the standard spectral function (see Eq.~(4) of the letter), 
we find that particle number $\mathcal{N}(t)$ and energy $\mathcal{E}(t)$ 
(see Eqs.~(7) and (8) of the letter, respectively) reduce to the definitions
\begin{align}
N(t) =& \mbox{Tr} \bigg[ \hat d^\dagger\, \hat d\, \hat \rho(t) \bigg]
\\
\label{ENSan}
E(t) =& \mbox{Tr} \bigg[\bigg(\hat H_S(t) + \sum_{\nu} \hat V_{\nu}(t)/2\bigg) \hat \rho(t) \bigg].\end{align}
Also, if $\mathcal{C}_{\nu}(t,E)$ (see Eq.~(5) of the letter) is replaced (by hand) by 
\begin{align}
C_\nu(t,E) =& \mathcal{C}_{\nu}(t,E)+\{\Lambda_\nu(t,E);A(t,E)\,\phi(t,E)\} \nonumber \\
&+ \{\Gamma_\nu(t,E)\,\phi(t,E);\mbox{Re}\, G^r(t,E)\} \\
=& \{E-\varepsilon(t);A(t,E)\,\phi(t,E)\} 
\end{align}
in the expressions for the fluxes $\mathcal{I}_{\nu}(t)$ and $\mathcal{J}_{\nu}(t)$ 
(see Eqs.~(12) and (15) of the letter, respectively), one recovers the standard definition 
of the particle flux \cite{MeirWingreenPRL92, JauhoWingreenMeirPRB94, EspoLindVdBNJP10, MahanBook}
\begin{align}
\label{CurrNSan}
I_{\nu}(t) =& - \sum_{k\in {\nu}} \mbox{Tr} \big[ \hat c^\dagger_{k} \hat c_{k} d_t \hat \rho(t) \big],
\end{align}
and the heat flux definition 
\begin{align}
\label{CurrESan}
J_{\nu}(t) =& -\mbox{Tr} \bigg[ \bigg(\hat H_\nu+ \hat V_\nu(t)/2\bigg)d_t \hat \rho(t) \bigg] \\
&+ \frac{1}{2}\mbox{Tr} \bigg[ d_t \hat V_{\nu}(t) \hat \rho(t) \bigg] \nonumber,
\end{align}
which was proposed (without the last term) in Ref.~\cite{SanchezPRB14} for a single-level coupled 
to a single reservoir model with an external driving only acting on the position of the level. 
Eq.~(\ref{CurrESan}) is thus a generalization which also takes into account driving in the coupling.

We emphasize that these definitions have been recovered by simply disregarding without 
justifications terms that are essential to formulate a consistent nonequilibrium thermodynamics. 
Related considerations have been made within the framework of the quantum Boltzmann equation 
\cite{IvanovKnollVoskresenskyNuclPhysA99, KitaProgrTheorPhys10}).

We finally note that at steady-state, $C_\nu(t,E) = \mathcal{C}_{\nu}(t,E)$ is satisfied, 
so that the current expressions (\ref{CurrNSan}) become exact and (\ref{CurrESan}) further 
simplifies to $J_{\nu}(t)=-\mbox{Tr} [ \hat H_\nu d_t \hat \rho(t) ]$.

%%%%%%%%%%%%%%%%%%%%%%%%%%%%%%%%%%%%%%%%%%%%%%%%%%%%%%%%%%%%%%%%%%%%%%
%\bibliography{BibFile}

%%%%%%%%%%%%%%%%%%%%%%%%%%%%%%%%%%%%%%%%%%%%%%%%%%%%%%%%%%%%%%%%%%%%%%
\end{document}